\documentclass[pra,twocolumn,superscriptaddress,notitlepage,showpacs,showkeys]{revtex4-1}
\usepackage{graphicx,amsmath,amsfonts,amssymb,upgreek,txfonts,color}
\usepackage[colorlinks,linkcolor=blue,citecolor=blue,urlcolor=blue,breaklinks=true]{hyperref}
\usepackage[utf8x]{inputenc}

\begin{document}

\title{A Green's function approach to the linear response of a driven dissipative optomechanical system}

\author{Ali Motazedifard} 
\email{motazedifard.ali@gmail.com}
\address{Department of Physics, University of Isfahan, Hezar-Jerib, 81746-73441, Isfahan, Iran}
\address{Quantum Optics Group, Department of Physics, University of Isfahan, Hezar-Jerib, 81746-73441, Isfahan, Iran}
\address{Quantum Communication $ \& $ Quantum Optics group, Iranian Center for Quantum Technologies (ICQTs), Tehran, Iran}

\author{A. Dalafi} 
\email{a\_dalafi@sbu.ac.ir}
\address{Laser and Plasma Research Institute, Shahid Beheshti University, Tehran 19839-69411, Iran}

\author{M. H. Naderi} 
\email{mhnaderi@sci.ui.ac.ir}
\address{Department of Physics, University of Isfahan, Hezar-Jerib, 81746-73441, Isfahan, Iran}
\address{Quantum Optics Group, Department of Physics, University of Isfahan, Hezar-Jerib, 81746-73441, Isfahan, Iran}

\date{\today}
\begin{abstract}
In this paper, we first try to shed light on the ambiguities that exist in the literature in the generalization of the standard linear response theory (LRT) which has been basically formulated for closed systems to the theory of open quantum systems in the Heisenberg picture. Then, we investigate the linear response of a driven-dissipative optomechanical system (OMS) to a weak time-dependent perturbation using the so-called generalized LRT. It is shown how the Green's function equations of motion of a standard OMS as an open quantum system can be obtained from the quantum Langevin equations (QLEs) in the Heisenberg picture. The obtained results explain a wealth of phenomena, including the anti-resonance, normal mode splitting  and the optomechanically induced transparency (OMIT). Furthermore, the reason why the Stokes or anti-Stokes sidebands are amplified or attenuated in the red or blue detuning regimes is clearly explained which is in exact coincidence, especially in the weak-coupling regime, with the Raman-scattering picture.
\end{abstract}


\maketitle

\section{Introduction}
Recent developments in quantum science and quantum technologies have attracted much attention to more profound theoretical investigations of driven-dissipative quantum phenomena. In the recent two decades, optomechanical systems (OMSs) in which the vibrational mode of a mechanical oscillator is coupled to the electromagnetic mode of a cavity via the radiation pressure \cite{Aspelmeyer}, have been known as one of the best schemes to investigate such phenomena. Other examples include ultra cold atoms \cite{Maschler2008} and Bose-Einstien condensates \cite{Morsch,Brenn Nature, Brenn Science,Ritter Appl. Phys. B} trapped inside optical lattices emerging optomechanical properties\cite{dalafi1,dalafi2,dalafi3,dalafi4,dalafi5}, and solid state systems such as superconducting qubits \cite{Schoelkopf} or arrays of superconducting microwave cavities \cite{Wallraff, Houck, Hur}.

In many of the above-mentioned experimental setups, it is very straightforward and interesting to study the response of the system to a weak external time-dependent perturbation \cite{oMITReview2018,vitaliOMIT2}. Nevertheless, the theoretical modeling of such phenomena could be very challenging. Generally, the view point of the well-known linear response theory (LRT), as has been raised in the textbooks on the many body physics and condensed matter theory \cite{Coleman, Flensberg, Stefanucci}, is based on a closed model of the quantum system which is in contact with a thermal bath at a finite temperature and is driven by a weak external source. Although this approach can give us the system Green's functions and describe its linear response correctly, the effect of dissipation is entered into the theory phenomenologically.

On the other hand, the theory of open quantum systems \cite{Carmichael,Bowen book,Gardiner,zubairy} which is the most profound and realistic approach to the description of quantum systems interacting with their environments, is so complete that can describe any aspect of a quantum system without any necessity of such phenomenological manipulations. Nevertheless, there are elements of ambiguities which are not addressed in the literature when one tries to describe the linear response of the system in the framework of open quantum systems. In spite of its importance, only few references \cite{Arrigoni, Dorda, greenScarlatella1, greenScarlatella2,keldyshReview2016,banPRA17,banQStud15,shenPRA17} have dealt with the Green's functions of open quantum systems whose approaches are mainly based on the Lindblad master equation, i.e., presented in the Schr{\"o}dinger picture.

Motivated by the above-mentioned studies and in order to resolve the existing ambiguities, in this paper we firstly present a complete formulation of the so-called \textit{generalized} LRT which is based on an open model of quantum systems as a more sophisticated approach against the \textit{standard} LRT raised in the context of condensed matter theory. 
In this \textit{generalized} LRT, the linear response of each system variable to an external perturbation is obtained through its corresponding open quantum system Green's functions. Furthermore, we show that by using the method of Green's function equation of motion, not only the evolutions of the open quantum system Green's functions are derived but also the relations between the system susceptibilities and Green's functions are clarified. As a simple example, we illustrate the application of the \textit{generalized} LRT for a single-mode driven-dissipative quantum field.

Then, as a practical and realistic open quantum system, we investigate the linear response of a standard OMS, to a weak external time-dependent potential. It is shown that the OMS Green's functions satisfy a set of ordinary differential equations which can be derived through the quantum Langevin equations (QLEs) of the system variables. Interestingly, in the case of the linearized OMS investigated here, the equations of motion of the Green's functions have exact analytical solutions. Nevertheless, it should be noted that although the presented approach is systematically applicable for any kind of open quantum systems even with nonlinearity, the Green's function equations of motion do not necessarily have exact analytical solutions in general.  
 
Applying the \textit{generalized} LRT to a standard OMS in the linearized regime, we obtain both the optical and the mechanical responses of the OMS to a weak time-dependent perturbation which drives the cavity field. The obtained results give us a precise and complete description of a wealth of phenomena like the anti-resonance \cite{Belbasi,Joe}, normal mode splitting \cite{Dobr,Grob} and the optomechanically induced transparency (OMIT) \cite{OMIT1,OMIT2,OMIT3,vitaliOMIT,marquardtOMIT,XiongOMIT} which is analogous to the familiar phenomenon of the electromagnetically induced transparency (EIT) \cite{EIT}. 

Furthermore, the \textit{generalized} LRT explains how the anti-Stokes and Stokes sidebands are amplified, respectively, in the red- and blue-detuned regimes. The obtained results are in exact coincidence, especially in the weak-coupling regime, with the Raman-scattering picture \cite{Wilson} which demonstrates how the optical shot noise affects the transition rates between the energy levels of the OMS. As an important result, the contributions of the beam-splitter (BS) as well as the two-mode squeezing (TMS) interactions on the Stokes and anti-Stokes sidebands are studied in the presence and absence of the rotating wave approximation (RWA).

The other important outcome of the generalized LRT which has been investigated in this paper is the precise description of the phenomenon of anti-resonance in a standard OMS. The anti-resonance \cite{Belbasi,Joe} occurs in every dynamical system (no matter being quantum or classical) consisting of two or several coupled oscillators. Specifically, in the case of two coupled oscillators, like the standard OMS, there is one anti-resonance frequency just for the oscillator which is directly driven by the external source (the optical field in OMS) where its amplitude of oscillation goes to zero.


The paper is organized as follows: In Sec.\ref{sec GLRT} we review the generalization of the LRT to the open quantum systems in the Heisenberg picture and apply it for a single-mode open quantum field. In Sec.\ref{secOMS} the Hamiltonian of the standard OMS is introduced in the linearized regime and the QLEs are derived. In Sec.\ref{sec LR OMS} the \textit{generalized} LRT is applied for the standard OMS and the optical and the mechanical responses of the OMS are obtained based on the open system Green's function. Finally, the summary and conclusions are given in Sec.\ref{sec conclusion}.

\section{Generalization of the Linear Response Theory to the Open Quantum Systems \label{sec GLRT}}
As was mentioned in the Introduction, there are elements of ambiguities in the literature in the description of the linear response of an open quantum system. Here, we are going to shed light on these ambiguities which have not been addressed with such details in the literature. For this purpose, we firstly present a complete formulation in which the LRT is incorporated into the theory of open quantum systems in the Heisenberg picture in spite of Refs. \cite{Arrigoni, Dorda, greenScarlatella1, greenScarlatella2,keldyshReview2016,banPRA17,banQStud15,shenPRA17} where the LRT has been investigated in the Schrödinger picture. Our main goal is to show how one can derive the equations of motion for the Green's functions of an open quantum system through the QLEs corresponding to the open system operators. In the next two subsections, after presenting the formulation of the generalized LRT, we derive the Green's functions equations of motion for a single-mode open quantum field from its QLEs as a simple example of an open quantum system.
\subsection{Theoretical formalism} 

We are going to obtain the linear response of a quantum system with the Hamiltonian $\hat H_S$ to an external time-dependent perturbation $\hat V(t)$ in the framework of the theory of open quantum systems when the environment is described by a multi-mode quantum field \cite{Carmichael}. The total Hamiltonian of the open system in the presence of the perturbation $\hat V(t)$ is described by
\begin{equation}\label{Htot}
	\hat H=\hat H_0+\hat V(t),
\end{equation}
where $\hat H_0$ includes the free Hamiltonian of the system ($\hat H_S$), the Hamiltonian of the reservoir ($\hat H_R$), and the Hamiltonian of the system-reservoir interaction ($\hat H_{SR}$):
\begin{equation}\label{H0 open}
	\hat H_0=\hat H_S+\hat H_R+\hat H_{RS}.
\end{equation}

Let us assume that the perturbation is turned on at $t=t_0$. In order to obtain the response of an arbitrary operator $\hat u$ of the system to the external perturbation, we can obtain its time evolution in the Heisenberg picture as
\begin{equation}\label{AHt}
	\hat u_H(t)=\hat U_I^\dag (t,t_0) \hat u_I(t) \hat U_I (t,t_0),
\end{equation}
where
\begin{equation}\label{UT}
	\hat U_I (t,t_0)=T \exp\Bigg(-\frac{i}{\hbar}\int_{t_0}^{t} dt^{\prime} \hat V_{I}(t^{\prime}) \Bigg)
\end{equation}
is the time evolution operator in the interaction picture in which $T$ stands for the time ordering operator \cite{Coleman}. Besides, the operators in the interaction picture are defined as
\begin{subequations}
	\begin{eqnarray}
		&&\hat u_I(t)=e^{\frac{i}{\hbar}\hat H_0 (t-t_0)} \hat u(t_0) e^{-\frac{i}{\hbar}\hat H_0 (t-t_0)},\label{uIt}\\
		&&\hat V_I(t)=e^{\frac{i}{\hbar}\hat H_0 (t-t_0)} \hat V(t_0) e^{-\frac{i}{\hbar}\hat H_0 (t-t_0)}.\label{VI(t)}
	\end{eqnarray}
\end{subequations}
The important point that should be noted here is that in the absence of the perturbation $\hat V(t)$ the system operators in the interaction picture are just the same operators in the Heisenberg picture  which obey the QLEs in the framework of the open quantum systems.

If the perturbation is very weak, the time evolution operator of Eq.(\ref{UT}) can be expanded up to the first order in $\hat V_I(t)$ as
\begin{equation}\label{UIFO}
	\hat U_I (t,t_0)\approx 1-\frac{i}{\hbar}\int_{t_0}^{t}dt^{\prime} \hat V_I (t^{\prime}).
\end{equation}
On substituting Eq.(\ref{UIFO}) into Eq.(\ref{AHt}) the operator $\hat u_H(t)$ is obtained up to the linear order in perturbation,
\begin{equation}\label{uHtFO}
	\hat u_H (t)=\hat u_I (t)-\frac{i}{\hbar}\int_{t_0}^{t} dt^{\prime} [\hat u_I (t), \hat V_I (t^{\prime})].
\end{equation}
If the state of the system is $\hat\rho_0$ at $t=t_0$ just before the time-dependent perturbation is turned on (which can be generally the steady-state of the system in the absence of the perturbation), then the expectation value of Eq.(\ref{uHtFO}) can be written as
\begin{equation}\label{EAt}
	\langle\hat u (t)\rangle=\langle\hat u\rangle_0-\frac{i}{\hbar}\int_{t_0}^{t} dt^{\prime} \langle[\hat u(t), \hat V(t^{\prime})]\rangle_0,
\end{equation}
where the left-hand side is the expectation value of $\hat u$ in the Heisenberg picture (in the presence of the time-dependent perturbation) which is given by $\langle\hat u_H (t)\rangle={\rm tr} ~ \Big(\hat u_H(t)\hat \rho_0\Big) $ while the expectation values on the right-hand side which have been indicated by the subscript $0$ are calculated in the interaction picture (which is equivalent to the Heisenberg picture in the absence of the time-dependent perturbation), e.g, $\langle\hat u\rangle_0={\rm tr}~\Big(\hat u_I(t)\hat \rho_0\Big)$. It should be reminded that in Eq.(\ref{EAt}) we have omitted the indices indicating the Heisenberg and interaction picture for simplicity. 

In the following subsection we illustrate the application of the \textit{generalized} LRT for a simple example of a single-mode driven-dissipative quantum field. In the next section, we apply the theory for the linearized OMS.

\subsection{Linear response of a single-mode open quantum field}
As a simple example, consider a single-mode quantum field $\hat a$ as an open quantum system which is driven by an external time-dependent perturbation. Physically, it can be the electromagnetic field inside a standard optical cavity with the resonance frequency $\omega_0$ which is driven by an external pump (probe) laser with the frequency $\omega_{p}$ \cite{zubairy}. The total Hamiltonian of the open system can be written as $\hat H=\hat H_0+\hat V(t)$ where (for more details see appendix \ref{appA}) 
\begin{equation}
\hat H_S=\hbar\omega_0\hat a^\dag\hat a,
\end{equation}
and
\begin{equation}\label{Vt0}
\hat V(t)=\hbar\eta (\hat a e^{i\omega_{p}t}+\hat a^\dagger e^{-i\omega_{p}t}) ,
\end{equation}
with $\eta$ being the pump rate of the external probe laser. 

Now, by rewriting the time-dependent potential of Eq.(\ref{Vt0}) in the interaction picture according to Eq.(\ref{VI(t)}) and substituting it into Eq.(\ref{EAt}) the response of the the field operator $\hat u=\hat a$ to the time-dependent perturbation is obtained as
\begin{eqnarray}\label{atm0}
\langle\hat a (t)\rangle&=&\langle\hat a\rangle_0-i\eta\int_{t_0}^{t} dt^{\prime} \langle[\hat a(t), \hat a(t^{\prime})]\rangle_0 e^{i\omega_{p}t}\nonumber\\
&&-i\eta\int_{t_0}^{t} dt^{\prime} \langle[\hat a(t), \hat a^\dag(t^{\prime})]\rangle_0 e^{-i\omega_{p}t},
\end{eqnarray}
where $\langle\hat a\rangle_0$ is the mean value at the steady state of the system before the time-dependent perturbation is turned on. Besides, the field operators under the integrals are in the interaction picture which is just the Heisenberg picture in the absence of the time-dependent perturbation (the subscript $I$ has been omitted). In other words, the time-evolution of $\hat a_I (t)$ is obtained either by Eq.~(\ref{uIt}) with $u=a$ or equivalently by the QLE of Eq.~(\ref{adottA}) which is obtained by eliminating the equations of the reservoir modes as has been shown in the appendix \ref{appA}. Therefore, they satisfy the QLEs in the framework of the theory of open quantum systems \cite{zubairy, Gardiner}, i.e.,
\begin{subequations}
	\begin{eqnarray}
	\dot{\hat a}(t)=-(i\omega_0+\kappa/2)\hat a(t)+\sqrt{\kappa}\hat a_{in}(t),\label{adott}\\
	\dot{\hat a}^\dag (t)=(i\omega_0-\kappa/2)\hat a^\dag (t)+\sqrt{\kappa}\hat a_{in}^\dag(t),\label{adagdott}
	\end{eqnarray}
\end{subequations}
where $\kappa$ is the damping rate of the cavity and $\hat a_{in}(t)$ is the input quantum noise which enters the cavity through the environment. It is obvious from Eq.(\ref{adott}) that the steady-state value of the field operator $\hat a$ is zero, i.e.,  $\langle\hat a\rangle_0=0$.

On the other hand, by defining the retarded Green's functions of the system as \cite{Coleman}
\begin{subequations}
	\begin{eqnarray}
	G_{aa}^R(t-t^{\prime})=-i\theta(t-t^\prime)\langle [\hat a(t),\hat a(t^{\prime})]\rangle_0,\label{Gaa0}\\
	G_{aa^\dag}^R(t-t^{\prime})=-i\theta(t-t^\prime)\langle [\hat a(t),\hat a^\dag(t^{\prime})]\label{Gaad0}\rangle_0,
	\end{eqnarray}
\end{subequations}
where $\theta(t-t^{\prime})$ is the Heaviside step function, Eq.(\ref{atm0}) can be rewritten as
\begin{eqnarray}\label{LRT at0}
\langle\hat a(t)\rangle=\langle\hat a\rangle_0 + \eta\int_{-\infty}^{+\infty} dt^{\prime} G_{aa}^R(t-t^{\prime}) &&e^{i\omega_{p}t^{\prime}}\nonumber\\
+\eta\int_{-\infty}^{+\infty} dt^{\prime} G_{aa^\dag}^R(t-t^{\prime}) e^{-i\omega_{p}t^{\prime}},
\end{eqnarray}

The important point is that the system Green's functions can be obtained from the field equations of motion in the Heisenberg picture, i.e., QLEs (\ref{adott}) and (\ref{adagdott}). 

In order to obtain the equations of motion for the Green's function $G_{aa}^R(\tau)$  $\big(G_{aa^\dag}^R(\tau)\big)$ with $\tau=t-t^\prime$, one should multiply Eq.(\ref{adott}) by $\hat a(0)$ $\big(\hat a^\dag(0)\big)$ on the left and on the right, subtract them from each other and then take their mean values. In this way, the equations of motion for the Green's functions are obtained as
\begin{subequations}
	\begin{eqnarray}
	\frac{d}{d\tau}G_{aa}^R(\tau)&=&-(i\omega_0+\kappa/2)G_{aa}^R(\tau),\label{Gaatau}\\
	\frac{d}{d\tau}G_{aa^\dag}^R(\tau)&=&-i\delta(\tau)-(i\omega_0+\kappa/2)G_{aa^\dag}^R(\tau).\label{Gaadagtau}
	\end{eqnarray}
\end{subequations}

On the other hand, it is easy to show that Eq.(\ref{LRT at0}) can be written in terms of the Fourier transforms of the Green's functions, i,e., $ \tilde G(\omega)= \int_{-\infty}^{\infty}d\tau G(\tau)  e^{i\omega \tau} $, as follows
\begin{equation}\label{lrsmf}
\langle\hat a(t)\rangle=\eta \tilde G_{aa^\dag}^R(\omega_{p}) e^{-i\omega_{p}t}+\eta \tilde G_{aa}^R(-\omega_{p}) e^{i\omega_{p}t}.
\end{equation}
Now, by taking the Fourier transforms of the set of differential equations (\ref{Gaatau}) and (\ref{Gaadagtau}) it can be easily shown that
\begin{subequations}
	\begin{eqnarray}
	\tilde G_{aa}^R(\omega)&=&\delta(\omega-\omega_0+i\kappa/2),\label{Gaawp}\\
	\tilde G_{aa^\dag}^R(\omega)&=&\frac{1}{\omega-\omega_0+i\kappa/2}.\label{Gaadagwp}
	\end{eqnarray}
\end{subequations}
In the region around the resonance frequency of the cavity where $\omega_{p}>0$, it is obvious that $\tilde G_{aa}^R(-\omega_{p})=0$ according to Eq.~(\ref{Gaawp}). Therefore, the response of the field operator to the external perturbation, i.e., Eq.~(\ref{lrsmf}) reads as follows
\begin{equation}\label{LRS empcav}
\langle\hat a(t)\rangle=\frac{\eta}{\omega_{p}-\omega_0+i\kappa/2} e^{-i\omega_{p}t},
\end{equation}
which is exactly what is expected as the mean value of the operator $\hat a$ when the oscillator is driven by an external source with the frequency $\omega_{p}$. As is seen, the amplitude of oscillation is just the Green's function of Eq.(\ref{Gaadagwp}) which is a Lorentzian function of $\omega_{p}$ with a single peak at $\omega_{p}=\omega_0$ with a width of $\kappa/2$. The important point is that the damping rate of the system has been appeared in the Green's function without any phenomenological manipulation and has been manifested as a natural consequence of the \textit{generalized} LRT.

\section{Standard bare OMS \label{secOMS}}

\begin{figure}
	\includegraphics[width=7cm]{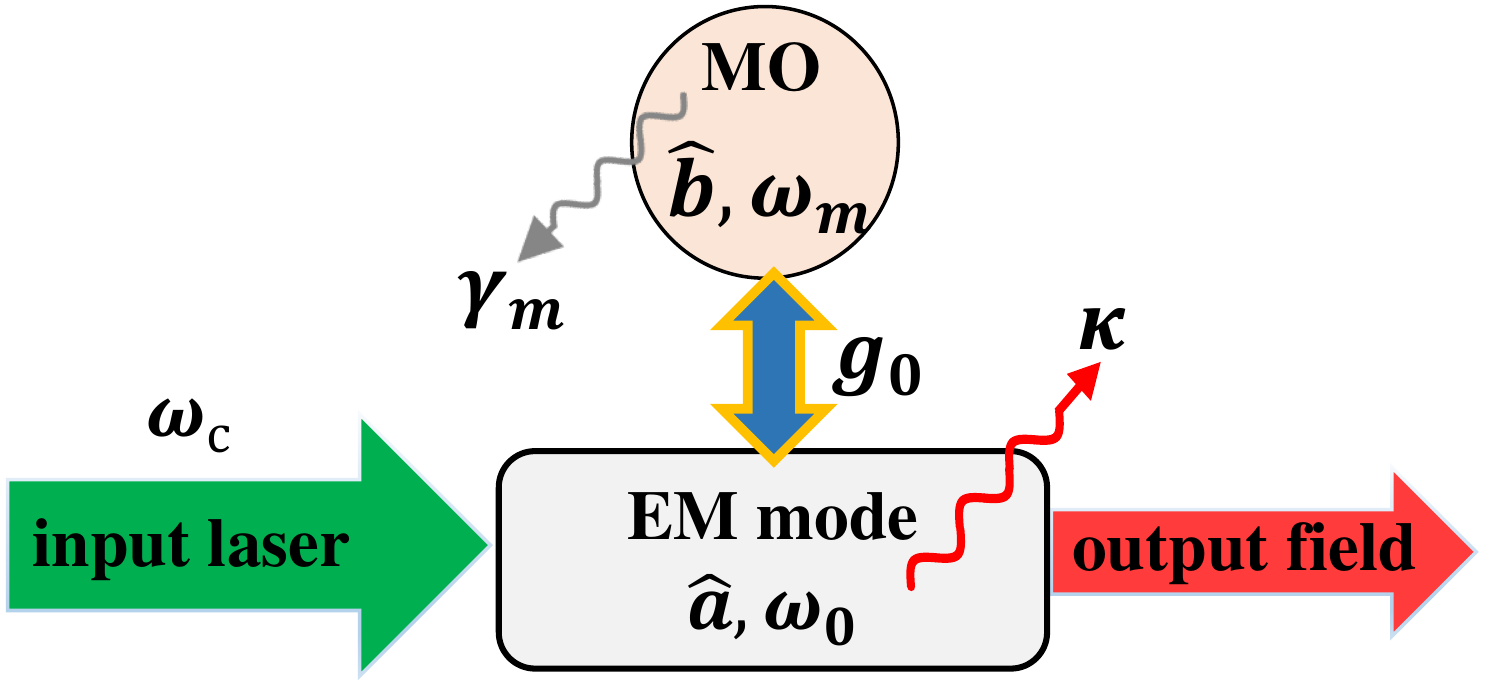}
	\caption{(Color online) Schematic of a general OMS in which a mechanical (phononic) mode $ \hat b $ is coupled to a classically driven quantized electromagnetic mode (EM) $ \hat a $ via the radiation pressure with strength $ g_0 $. The natural frequencies and dissipation rates of the EM mode and mechanics are respectively $ \omega_0,\kappa $ and $ \omega_m, \gamma_m $ while $ \omega_c $ being the laser frequency.}
	\label{fig1}
\end{figure}

As has been shown schematically in Fig.~(\ref{fig1}), we consider a standard OMS in which the cavity field with the resonance frequency $\omega_0$ is coupled to a mechanical oscillator (MO) with the natural frequency $\omega_{m}$. The cavity field is pumped by an external (coupling) laser with frequency $\omega_{c}$ at the rate of $\eta$. The dynamics of the system is governed by the nonlinear Hamiltonian \cite{Aspelmeyer}
\begin{eqnarray}\label{NL}
\hat H_{OMS}&=&\hbar\omega_0\hat a^\dag \hat a+\hbar\omega_{m}\hat b^\dag \hat b+\hbar g_0 \hat a^\dag \hat a(\hat b+\hat b^\dag)\nonumber\\
&&+\hbar\eta(\hat a e^{i\omega_c t}+\hat a^\dag e^{-i\omega_c t}),
\end{eqnarray}
where $ \hat a (\hat b) $ is the annihilation operator of the optical (mechanical) mode and $g_0$ is the single-photon optomechanical coupling. The nonlinear Hamiltonian of Eq.(\ref{NL}) can be linearized by considering the operators as $\hat a=\alpha+\delta\hat a$ and $\hat b=\beta+\delta\hat b$ where $\alpha \approx -\eta/\Delta$ and $\beta\approx -g_0\alpha^2/\omega_{m}$ are, respectively, the mean fields of the optical and mechanical modes in the resolved-sideband regime and $\Delta=\omega_0 -\omega_c -2g^2/\omega_{m}$ is the effective cavity detuning with $g=g_0\alpha$ being the effective optomechanical coupling and $\delta\hat a$ ($\delta\hat b$) is the quantum fluctuation of the optical (mechanical) field around the mean-field value \cite{Bowen book}.

In the framework of open quantum systems and in the rotating frame oscillating at the coupling laser frequency $\omega_{c}$ the linearized Hamiltonian of the open system is given by \cite{Bowen book}
\begin{subequations}
\begin{eqnarray}
\hat H_{0}&=&\hat H_{S}+\hat H_{\kappa}+\hat H_{\gamma},\label{H0}\\
\hat H_S&=&\hbar \Delta \delta\hat a^\dag \delta\hat a + \hbar \omega_m \delta\hat b^\dag \delta\hat b + \hbar g_b (\delta\hat a \delta\hat b^\dag + \delta\hat a^\dag \delta\hat b)\label{HS}\nonumber\\
&&+\hbar g_t (\delta\hat a \delta\hat b + \delta\hat a^\dag \delta\hat b^\dag),
\end{eqnarray}	
\end{subequations}
where $\hat H_{S}$ is the linearized version of the nonlinear Hamiltonian of Eq.(\ref{NL}) and $\hat H_{\kappa}$ ($\hat H_{\gamma}$)is the Hamiltonian of the optical (mechanical) reservoir together with its interaction with the optical (mechanical) mode. Here, we have defined $g_b=g \epsilon_b$ to indicate the contribution of the beam-splitter (BS) interaction \cite{Aspelmeyer} with $\epsilon_b=1$ (the third term of Eq.(\ref{HS})) and also have used $g_t=g \epsilon_t$ to indicate the contribution of the two-mode squeezing (TMS) interaction \cite{Aspelmeyer} with $\epsilon_t=1$ (the fourth term of Eq.(\ref{HS})). In this way, in the red-detuned regime of $\Delta=+\omega_m$ where the TMS interaction is suppressed one can easily take $\epsilon_t=0$ while in the blue-detuned regime of $\Delta=-\omega_m$ where the BS interaction is suppressed one can take $\epsilon_b=0$ in the rotating wave approximation (RWA).

The QLEs corresponding to the linearized Hamiltonian introduced by Eqs.(\ref{H0}) and (\ref{HS}) can be written as
\begin{subequations}
	\begin{eqnarray}
	&& \delta\dot{\hat a} = -(\kappa/2+ i \Delta) \delta\hat a -i g_b \delta\hat b -i g_t \delta\hat b^\dag + \sqrt{\kappa} \hat a_{in}, \label{equOMSa} \\
	&& \delta\dot {\hat b} = -(\gamma_m/2+ i \omega_m) \delta\hat b -i g_b \delta\hat a -i g_t \delta\hat a^\dag + \sqrt{\gamma_m} \hat b_{in}. \label{equOMSb}
	\end{eqnarray}
\end{subequations}
The linearized QLEs of motion together with their Hermitian conjugates can be rewritten in the following compact form
\begin{eqnarray} \label{u1}
\frac{d}{dt} \hat{ \boldsymbol{u}}(t)= \boldsymbol{\chi}_0 \hat{ \boldsymbol{u}}(t) + \hat{ \boldsymbol{u}}_{in}(t), 
\end{eqnarray}
where $ \hat{ \boldsymbol{u}}(t)=\Big(\delta\hat a(t), \delta\hat a^\dagger(t), \delta\hat b(t), \delta\hat b^\dagger(t)\Big)^{\rm T} $ is the vector of the quantum fields fluctuations, and $  \hat{ \boldsymbol{u}}_{in}(t)= \Big(\sqrt{\kappa} \hat a_{in},\sqrt{\kappa} \hat a^{\dag}_{in},\sqrt{\gamma_m} \hat b_{in}, \sqrt{\gamma_m} \hat b^{\dag}_{in} \Big)^{\rm T} $ is the vector of quantum noises. The drift matrix $ \boldsymbol{\chi}_0  $ can be found easily from the equations of motion as follows
\begin{eqnarray} \label{chi0OMS}
&&\!\!\!\!\!\!\!\!\!\!\!\!  \boldsymbol{\chi}_0\!=\! \left( \begin{matrix}
{-\frac{\kappa}{2}-i\Delta} & {0} & {-ig_b} & {-ig_t }  \\
{0} & {-\frac{\kappa}{2}+i\Delta} & {ig_t } & {ig_b } \\
{-ig_b } & {-ig_t } & {-\frac{\gamma_m}{2}-i\omega_m} & {0} \\
{ig_t } & {ig_b }& {0} & {-\frac{\gamma_m}{2}+i\omega_m}  \\
\end{matrix} \right).
\end{eqnarray}

It is obvious that one can solve Eq.(\ref{u1}) in the Fourier space as follows
\begin{eqnarray} \label{u_w}
&& \hat{ \boldsymbol{u}}(\omega)= \boldsymbol{\chi}(\omega) \hat{ \boldsymbol{u}}_{in}(\omega), 
\end{eqnarray}
where the susceptibility matrix $  \boldsymbol{\chi}(\omega)  $ is generally defined as
\begin{eqnarray}\label{chiw}
&&  \boldsymbol{\chi}(\omega) = \Big(-i\omega\boldsymbol{1} -  \boldsymbol{\chi}_0 \Big)^{-1},
\end{eqnarray}
in which $\boldsymbol{1}$ is the $4\times 4$ idendity matrix. Solving Eq.(\ref{u_w}) for the mechanical mode, one can obtain the following equations
\begin{subequations}
\begin{eqnarray} \label{bsolveOMS1}
&& \!\!\!\!\!\!\!\! \delta\hat b(\omega)= \chi_{-m}(\omega) \Big(-ig_b \delta\hat a(\omega) - i g_t \delta\hat a^\dag(\omega)+ \sqrt{\gamma_m} \hat b_{in}(\omega)\Big),\label{bwso} \\
&& \!\!\!\!\!\!\!\! \delta\hat b^\dag(\omega)= \chi_{+m}(\omega) \Big(ig_t \delta\hat a(\omega) + i g_b \delta\hat a^\dag(\omega)+ \sqrt{\gamma_m} \hat b_{in}^\dag(\omega)\Big)\label{bdagwso},
\end{eqnarray}	
\end{subequations}
where
\begin{equation}\label{Xpmm}
\chi_{\pm m}(\omega)= [\gamma_m/2 - i (\omega\pm \omega_m)]^{-1}.  \\
\end{equation}
Now, by substituting Eqs.(\ref{bwso}) and (\ref{bdagwso}) in the equations corresponding to the optical field, the following set of equations will be obtained for the optical mode 
\begin{subequations}
	\begin{eqnarray}
	-i\omega \delta\hat a(\omega)&=& \! -\Big(\frac{\kappa}{2} + i\Sigma_a(\omega)\Big) \delta\hat a(\omega) \nonumber\\
	&&+ \lambda_a (\omega) \delta\hat a^\dag (\omega) + \!\! \sqrt{\kappa} \hat A_{in}(\omega),\label{asolveOMS1 a}\\
	-i\omega \delta\hat a^\dag(\omega)&=&\! -\Big(\frac{\kappa}{2} -i\Sigma_a^\ast(-\omega)\Big) \delta\hat a^\dag (\omega)\nonumber\\
	&&+ \lambda_a^\ast(-\omega) \delta\hat a(\omega) + \!\! \sqrt{\kappa} \hat A_{in}^\dag(\omega),\label{asolveOMS1 b}
	\end{eqnarray}
\end{subequations}
which are very similar to those of a degenerate optical parametric amplifier (DPA) \cite{optomechanicswithtwophonondriving}. Here
\begin{equation}\label{lambdaa}
\lambda_a(\omega)= g_b g_t [\chi_{+m}(\omega)-\chi_{-m}(\omega)], \\
\end{equation}
plays the role of an effective modulation parameter \cite{aliDCE3} or an effective induced cavity squeezing coefficient which is also analogous to the parametric amplification in OPA. Moreover,
\begin{equation}
\Sigma_a(\omega)=\Delta -i g^2_b\chi_{-m}(\omega)+ig^2_t\chi_{+m}(\omega). \\
\end{equation}
is the effective cavity self-energy \cite{Aspelmeyer}. Furthermore, the last term in Eq.(\ref{asolveOMS1 a}) is an effective quantum noise which is given by
\begin{equation} \label{chiOMS}
\hat A_{in}(\omega)= \hat a_{in}(\omega)-i\sqrt{\frac{\gamma_m}{\kappa}}\Big[g_b\chi_{-m}(\omega)\hat b_{in}(\omega) + g_t\chi_{+m}(\omega) \hat b^\dag_{in}(\omega)\Big].
\end{equation}
As is seen from Eqs.(\ref{lambdaa}) and (\ref{Xpmm}), $ \lambda_a^\ast(-\omega)=-\lambda_a(\omega) $ and also $ \chi^\ast_{\pm m}(-\omega)= \chi_{\mp m}(\omega)$.
Finally, by solving the set of Eqs.(\ref{asolveOMS1 a}) and (\ref{asolveOMS1 b}) the optical field fluctuation $\delta \hat a(\omega)$ is obtained in the form of Eq.(\ref{u_w}) as
\begin{eqnarray} \label{a_w}
\delta\hat a(\omega)= && \chi_{aa}(\omega)\sqrt{\kappa}\hat a_{in}(\omega) + \chi_{aa^\dag}(\omega) \sqrt{\kappa}\hat a^\dag_{in}(\omega)\nonumber\\
&&+\chi_{ab}(\omega)\sqrt{\gamma_m}\hat b_{in}(\omega) + \chi_{ab^\dag}(\omega) \sqrt{\gamma_m}\hat b^\dag_{in}(\omega),
\end{eqnarray}
where the matrix elements of the susceptibility matrix $\boldsymbol{\chi}(\omega)$ are given by
\begin{subequations}
	\begin{eqnarray}
		&&\chi_{aa}(\omega)=\frac{Q^\ast (-\omega)}{D(\omega)},\label{Xaa}\\
	&&\chi_{aa^\dag}(\omega)=\frac{\lambda_a(\omega)}{D(\omega)},\label{Xaad}\\
	&&\chi_{ab}(\omega)=\frac{i \chi_{-m}(\omega)}{D(\omega)} \Big(g_t\lambda_a(\omega)-g_b Q^\ast(-\omega)\Big),\\
	&&\chi_{ab^\dag}(\omega)=\frac{i \chi_{+m}(\omega)}{D(\omega)} \Big(g_b\lambda_a(\omega)-g_t Q^\ast(-\omega)\Big),\label{Xabd}
	\end{eqnarray}
\end{subequations}
in which the functions $Q(\omega)$ and $D(\omega)$ have been defined as
\begin{subequations}
	\begin{eqnarray}
	&&Q(\omega)=\frac{\kappa}{2}-i\Big(\omega-\Sigma_a(\omega)\Big),\\
	&&D(\omega)=Q(\omega)Q^\ast(-\omega)-\lambda_a(\omega)\lambda_a^\ast(-\omega).
	\end{eqnarray}
\end{subequations}

Interestingly, as is seen from Eq.(\ref{lambdaa}) the effective parametric modulation in a standard optomechanical system exists only as far as both the BS and the TMS interactions are considered in the system Hamiltonian. Therefore, in the RWA where one of the mentioned inteactions is ignored the effective parametric modulation is vanished.

\section{linear response of the OMS to a weak external perturbation \label{sec LR OMS}}
In order to see how an OMS responds to a weak external time-dependent perturbation, assume that the linearized OMS described by the Hamiltonian of Eq.(\ref{HS}) is driven by a weak probe laser with frequency $\omega_p$ whose interaction with the system is described by the following time-dependent perturbation in the frame rotating at the coupling laser frequency
\begin{equation}\label{Vt}
\hat V(t)=\hbar\zeta\delta\hat a e^{i\omega_{pc}t}+\hbar\zeta^\ast\delta\hat a^\dagger e^{-i\omega_{pc}t},
\end{equation}
where $\omega_{pc}=\omega_{p}-\omega_{c}$ is the detuning between the probe and coupling lasers frequencies and $|\zeta|\ll\eta$. Therefore the OMS is described as an open quantum system by the following time-dependent Hamiltonian
\begin{equation}\label{HH0V}
\hat H=\hat H_0+\hat V(t),
\end{equation}
with $\hat H_0$ given by Eq.(\ref{H0}). Based on Eq.(\ref{VI(t)}) the time-dependent perturbation of Eq.(\ref{Vt}) in the interaction picture takes the form
\begin{equation}\label{VIt}
	\hat V_I(t)=\hbar\zeta\delta\hat a_I(t) e^{i\omega_{pc}t}+\hbar\zeta^\ast\delta\hat a_I^\dagger(t) e^{-i\omega_{pc}t},
\end{equation}
where $\delta\hat a_I(t)$ and its conjugate are in the interaction picture which is just the Heisenberg picture in the absence of the perturbation. In other words, $\delta\hat a_I(t)$ and its conjugate obey the QLEs given by Eqs.(\ref{equOMSa},\ref{equOMSb}) and their conjugates. In the following two subsections we investigate the linear responses of the optical and mechanical modes to the external time-dependent perturbation while we have deleted the subscript $I$ form the optical and mechanical modes.

\subsection{Optical response of the OMS}
Based on the \textit{generalized} LRT explained in Sec.\ref{sec GLRT}, the response of the optical field fluctuation $\delta\hat a(t)$ to the external time-dependent perturbation ($\ref{VIt}$) is obtained as follows by substituting Eq.(\ref{VIt}) into Eq.(\ref{EAt})
\begin{eqnarray}\label{LRT at}
\langle\delta\hat a(t)\rangle=\langle\delta\hat a\rangle_0+\zeta\int_{-\infty}^{+\infty} dt^{\prime} G_{aa}^R(t-t^{\prime}) &&e^{i\omega_{pc}t^{\prime}}\nonumber\\
+\zeta^\ast\int_{-\infty}^{+\infty} dt^{\prime} G_{aa^\dag}^R(t-t^{\prime}) e^{-i\omega_{pc}t^{\prime}},
\end{eqnarray}
in which $\delta\hat a$ has been substituted for $\hat u$. Based on Eq.(\ref{equOMSa}), $\langle\delta\hat a\rangle_0=0$ is the steady-state mean value of the optical field fluctuation in the absence of the time-dependent perturbation and
\begin{subequations}
	\begin{eqnarray}
	G_{aa}^R(t)=-i\theta(t)\langle [\delta\hat a(t),\delta\hat a(0)]\rangle_0,\label{Gaa}\\
	G_{aa^\dag}^R(t)=-i\theta(t)\langle [\delta\hat a(t),\delta\hat a^\dag(0)]\rangle_0\label{Gaad},
	\end{eqnarray}
\end{subequations}
are the cavity retarded Green's functions (CRGFs) of the linearized OMS described as an open quantum system by Eqs.(\ref{H0}-\ref{HS}). It should be again emphasized that the quantum field fluctuations in the Green's function definitions of Eqs.(\ref{Gaa}-\ref{Gaad}) should be considered in the interaction picture based on the Hamiltonian of Eq.(\ref{HH0V}) which is equivalent to the Heisenberg picture in the absence of the time-dependent perturbation $\hat V(t)$ [Sec.\ref{sec GLRT}]. In other words, the time evolutions of $\delta\hat a(t)$ and its conjugate are obtained from the QLEs derived in the previous section and the expectation values with subscript $0$ are calculated in the steady state of the system in the absence of the perturbation. It can be easily shown that Eq.(\ref{LRT at}) can be written as
\begin{equation}\label{mat}
\langle\delta\hat a(t)\rangle=\zeta^\ast \tilde G_{aa^\dag}^R(\omega_{pc}) e^{-i\omega_{pc}t}+\zeta \tilde G_{aa}^R(-\omega_{pc}) e^{i\omega_{pc}t}.
\end{equation}
where $\tilde G_{aa^\dag}^R(\omega_{pc})$ and $\tilde G_{aa}^R(-\omega_{pc})$ are the Fourier transforms of the Green's functions at $\omega=\pm\omega_{pc}$. Since $\langle\hat a(t)\rangle=\alpha+\langle\delta\hat a(t)\rangle$  with $\alpha\approx-\eta/\Delta$ being the steady-state optical mean field, the response of the optical field $\hat a(t)$ to the time-dependent perturbation in the laboratory frame is obtained as
\begin{equation}\label{mat2}
\langle\hat a(t)\rangle \!=\! \alpha e^{-i\omega_{c}t} \! +\zeta^\ast \tilde G_{aa^\dag}^R(\omega_{pc}) e^{-i(\omega_{c}+\omega_{pc})t}+\zeta \tilde G_{aa}^R(-\omega_{pc}) e^{-i(\omega_{c}-\omega_{pc})t}.
\end{equation}
As is seen from Eq.(\ref{mat2}) there is a central band oscillating with $\omega_{c}$ and two sidebands, the so-called Stokes and anti-Stokes sidebands, oscillating with $\omega_{c}\pm\omega_{pc}$ . It should be noted that, for $\omega_{pc}>0$ the second and the third terms  in the Eq.(\ref{mat2}) are, respectively, the anti-Stokes and the Stokes sidebands while for $\omega_{pc}<0$ they correspond, respectively, to the Stokes and the anti-Stokes sidebands. Therefore, the amplitudes of the anti-Stokes and Stokes sidebands
can be defined as follows
\begin{subequations}
	\begin{eqnarray}
	&& D_{AS}(\omega_{pc})=|\tilde G^{R}_{aa^\dag}(\omega_{pc})| \theta(\omega_{pc}) + |\tilde G^{R}_{aa}(-\omega_{pc})| \theta(-\omega_{pc}),\quad \label{DAS}\\
	&& D_{S}(\omega_{pc})=|\tilde G^{R}_{aa^\dag}(\omega_{pc})| \theta(-\omega_{pc}) + |\tilde G^{R}_{aa}(-\omega_{pc})| \theta(\omega_{pc}).\label{DS}
	\end{eqnarray}
\end{subequations}
where $\theta(\omega_{pc})$ is the Heaviside step function which is unity for $\omega_{pc}>0$ and is zero for $\omega_{pc}<0$.

In the following, we will calculate the CRGFs $G_{aa^\dag}^R(t)$ and $G_{aa}^R(t)$. For this purpose, we follow the approach of Ref.\cite{greenZoubi} in which Green's functions are obtained through a set of ordinary differential equations, the so-called Green's functions equations of motion, with the difference that we consider our system as an open quantum system while Ref.\cite{greenZoubi} is based on a closed model of the quantum system and the effects of dissipation has been fed into the equations phenomenologically. It should also be emphasized that the CRGFs of the linearized OMS obtained by the present approach is in complete coincidence with the approach of diagonalization of the Hamiltonian in terms of normal modes investigated in Ref.\cite{greenPRLclerk}.

In order to obtain the equations of motion for  $G_{aa^\dag}^R(t)$  $\big(G_{aa}^R(t)\big)$, one should multiply each of the equations in Eq.(\ref{u1}) by $\hat a^{\dagger}(0)$ $\big(\hat a(0)\big)$ on the left and on the right, subtract them from each other and then taking their mean values. In this way, the Green's functions equations of motion can be obtained as the following compact forms
\begin{subequations}
\begin{eqnarray}
&& \frac{d}{dt} \boldsymbol{G}^R_{a^\dag}(t)= -i\delta(t)\boldsymbol{V}_{a^\dag}+\boldsymbol{\chi}_0 \boldsymbol{G}^R_{a^\dag}(t),\label{GAd}\\
&& \frac{d}{dt} \boldsymbol{G}^R_{a}(t)= +i\delta(t)\boldsymbol{V}_{a}+\boldsymbol{\chi}_0 \boldsymbol{G}^R_{a}(t).\label{GA}
\end{eqnarray}
\end{subequations}
Here, $ \boldsymbol{G}^R_{a^\dag}(t)= \Big(G^R_{aa^\dag}(t), G^R_{a^\dag a^\dag}(t), G^R_{ba^\dag}(t), G^R_{b^\dag a^\dag}(t)\Big)^{\rm T} $, $ \boldsymbol{G}^R_{a}(t)= \Big(G^R_{aa}(t), G^R_{a^\dag a}(t), G^R_{ba}(t), G^R_{b^\dag a}(t)\Big)^{\rm T} $, and $ \boldsymbol{V}_{a^\dag}:= (1,0,0,0)^{\rm T} $ and $ \boldsymbol{V}_{a}:= (0,1,0,0)^{\rm T} $ are fixed four-dimensional vectors. Now, by taking the Fourier transforms of Eqs.(\ref{GAd}) and (\ref{GA}) one can find the Green's function vectors in the Fourier space as
\begin{subequations}
	\begin{eqnarray}
	&&  \boldsymbol{\tilde G}^R_{a^\dag}(\omega)= -i \boldsymbol{\chi}(\omega) \boldsymbol{V}_{a^\dag},\label{GAdw}\\
	&&  \boldsymbol{\tilde G}^R_{a}(\omega)= +i \boldsymbol{\chi}(\omega) \boldsymbol{V}_{a},\label{GAw},
	\end{eqnarray}
\end{subequations}
where $\boldsymbol{\chi(\omega)}$ is the susceptibility matrix defined by Eq.~(\ref{chiw}). As is seen from Eqs.~(\ref{GAdw}) and (\ref{GAw}), the CRGFs in the frequency space, i.e., the Fourier transform of Eqs.~(\ref{Gaa}) and (\ref{Gaad}) can be obtained as
\begin{subequations}
	\begin{eqnarray} \label{greenparametric}
	&& \tilde G^R_{aa^\dag}(\omega)=- i\chi_{aa} (\omega),\label{Gaad(w)}\\
	&& \tilde G^R_{aa}(\omega)=+ i\chi_{aa^\dag} (\omega).  \label{Gaa(w)}
	\end{eqnarray}
\end{subequations}
Based on Eqs.~(\ref{Gaad(w)}-\ref{Gaa(w)}), the absolute values of the anti-Stokes and Stokes amplitudes can be determined by $|\chi_{aa} (\omega)|$ and $|\chi_{aa^\dag} (-\omega)|$ depending on the sign of $\omega=\omega_{pc}$ which can be calculated straightforwardly through the set of Eqs.~(\ref{Xaa},\ref{Xaad}). 

On the other hand, the cavity photon spectral function (CPSF) is defined as
\begin{equation}
\rho(\omega)=-\frac{2}{\pi} {\rm{Im}} \tilde G_{aa^\dag}^R(\omega),
\end{equation}
which is usually interpreted as an effective density of single-particle states. In the well-known phenomenon of optomechanically induced transparency (OMIT) \cite{OMIT1,OMIT2,OMIT3,vitaliOMIT,marquardtOMIT,XiongOMIT} the suppression of the density of photon states, i.e., $\rho(\omega)$ at the cavity resonance leads to the perfect reflection of the weak probe field \cite{optomechanicswithtwophonondriving}. In OMIT, by injecting simultaneously a strong control driving laser and a weak probe laser with coherency times longer than the effective mechanical damping rate into the red-detuned sideband of an optomechanical system a cavity anti-resonance is provided which leads to the transparency of the weak probe laser \cite{Aspelmeyer,OMIT1,OMIT2,OMIT3,vitaliOMIT,marquardtOMIT,XiongOMIT}.

\begin{figure}
	\centering
	\includegraphics[width=4.25cm]{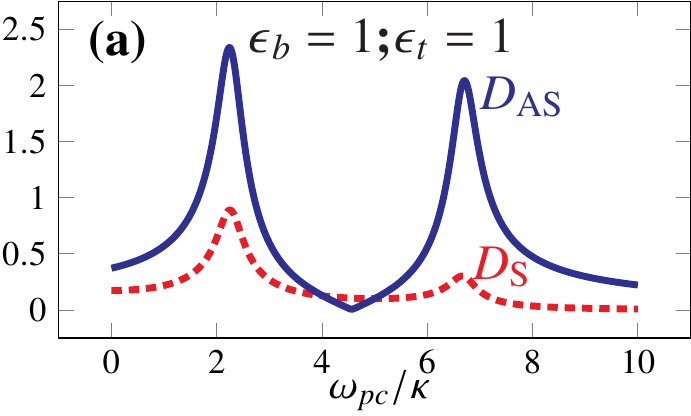}
	\includegraphics[width=4.25cm]{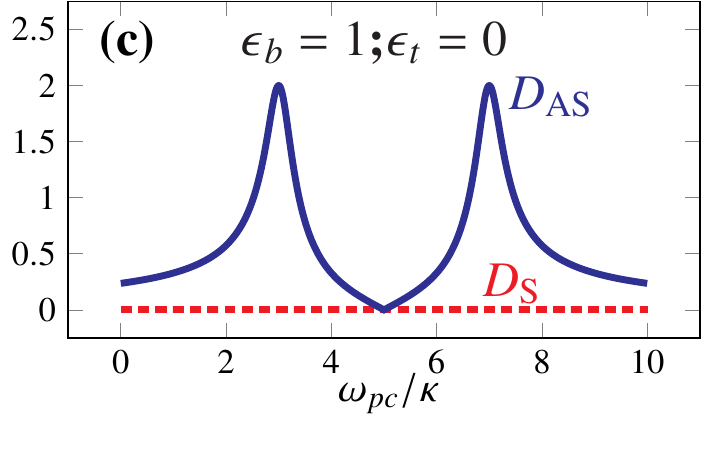}
	\includegraphics[width=4.25cm]{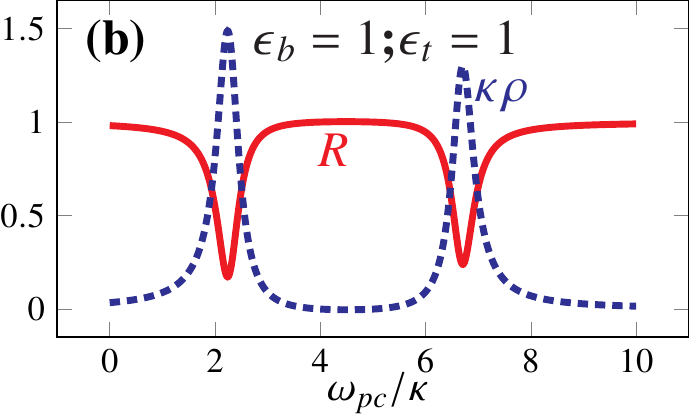}
	\includegraphics[width=4.25cm]{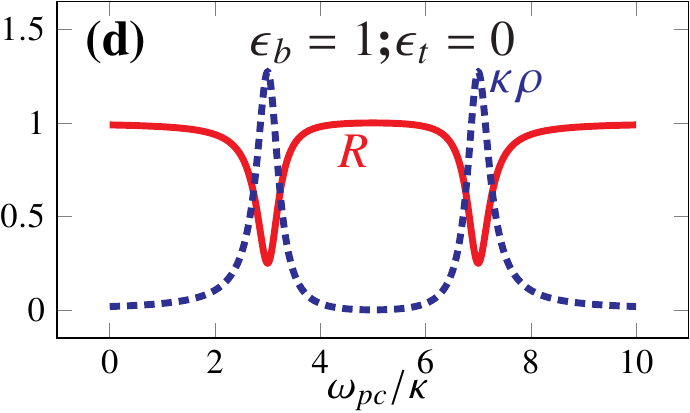}
	\caption{
		(Color online) (a),(c) The normalized amplitudes of the anti-Stokes $D_{AS}=\kappa|\tilde G^R_{aa^\dag}(\omega_{pc}/\kappa)|$ (blue solid curve) and the Stokes $D_{S}=\kappa |\tilde G^R_{aa}(-\omega_{pc}/\kappa)|$  (red dashed curve) sidebands of the optical field and (b), (d) the normalized spectral density function $\kappa\rho$ (blue dashed curve) and the reflection coefficient $R$ (red solid curve) versus the normalized frequency $\omega_{pc}/k$. Here, $(\epsilon_b=1,\epsilon_t=1)$ stands for the absence of the RWA and $(\epsilon_b=1,\epsilon_t=0)$ stand for the presence of the RWA. The system is in the red detuned regime of $\Delta=\omega_{m}$ with $g=2k$ The other parameters are: $\omega_{m}=5\kappa, \gamma_m=10^{-4}\kappa, \kappa_{cp}=0.25\kappa$.}
	\label{fig2}
\end{figure}


A standard linear response calculation shows that the elastic OMIT reflection coefficient is given by $r(\omega)=1-i \kappa_{cp} \tilde G_{aa^\dag}^R(\omega)$ where $\kappa_{cp}$ is the contribution to the total cavity damping rate $\kappa$ from the coupling to the drive port \cite{optomechanicswithtwophonondriving,greenPRLclerk}. It can be shown that the power reflection $R(\omega)=|r(\omega)|^2$ is approximately given by \cite{optomechanicswithtwophonondriving}
\begin{equation}\label{Rw}
R(\omega)\approx 1-\kappa_{cp}\rho(\omega).
\end{equation}

\begin{figure}[ht]
	\centering
	\includegraphics[width=4.25cm]{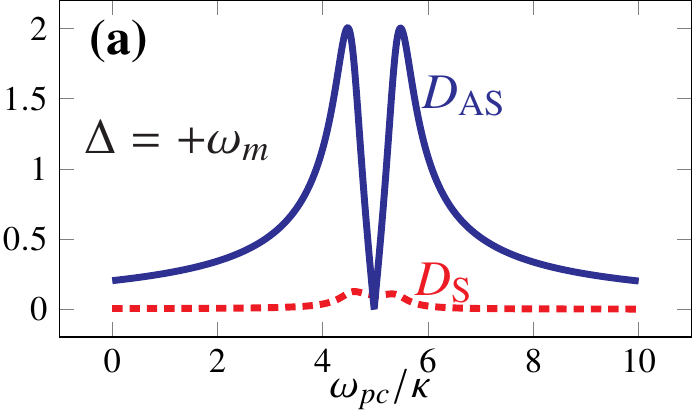}
	\includegraphics[width=4.25cm]{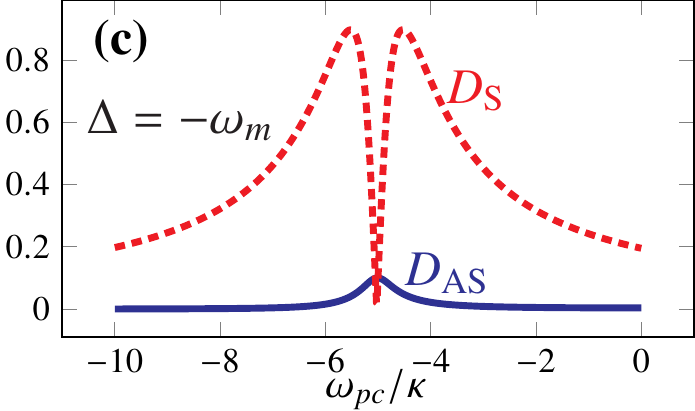}
	\includegraphics[width=4.25cm]{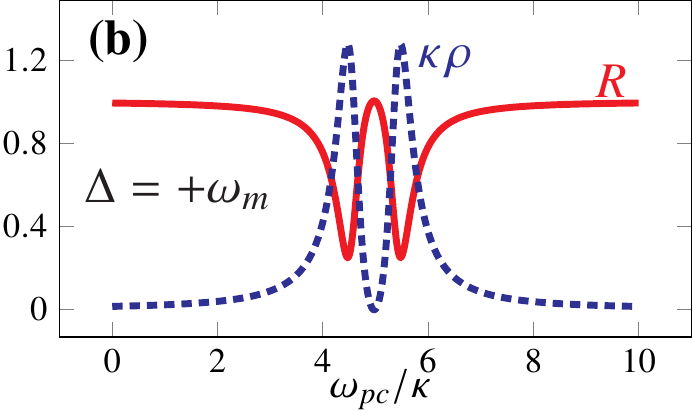}
	\includegraphics[width=4.25cm]{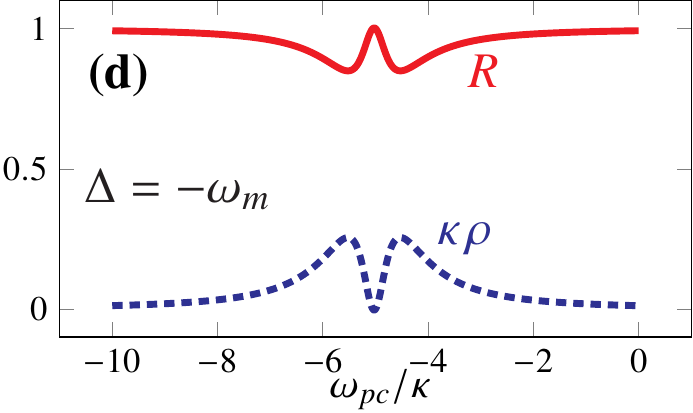}
	\caption{
		(Color online) (a),(c) The optical amplitudes of the anti-Stokes $D_{AS}$ (blue solid curve) and the Stokes $D_{S}$ (red dashed curve) sidebands and (b), (d) the normalized spectral density function $\kappa\rho$ (blue dashed curve) and the reflection coefficient $R$ (red solid curve) versus the normalized frequency $\omega_{pc}/\kappa$. Here, all the results have been obtained in the absence of the RWA $(\epsilon_b=1,\epsilon_t=1)$  with the optomechanical coupling of $g=0.5\kappa$. In (a) and (b) the system is in the red-detuned regime while in (c) and (d) it is in the blue-detuned regime. The other parameters are the same as those of Fig.(\ref{fig2}).}
	\label{fig3}
\end{figure}

As is seen from Eq.(\ref{Rw}), the power reflection is less than unity as far as the spectral function is positive. In OMIT the power reflection goes to unity near the cavity resonance because the spectral function is nearly zero at the cavity resonance. In the following we will show how all these phenomena can occur in a standard optomechanical system.

In order to show how the two kinds of interactions, i.e., the BS and TMS, can affect the linear response of the OMS, in Fig.(\ref{fig2}) we have plotted normalized amplitudes of the anti-Stokes $D_{AS}=\kappa|\tilde G^R_{aa^\dag}(\omega_{pc}/\kappa)|$ (blue solid curve) and the Stokes $D_{S}=\kappa |\tilde G^R_{aa}(-\omega_{pc}/\kappa)|$  (red dashed curve) sidebands of the optical field as well as the cavity spectral function (blue dashed curve) and the power reflection coefficient (red solid curve) versus the normalized frequency $\omega_{pc}/\kappa$ in the red-detuned regime where $\Delta=\omega_{m}$ while the other parameters have been considered as $g=2\kappa, \omega_{m}=5\kappa, \gamma_m=10^{-4}\kappa, \kappa_{cp}=0.25\kappa$.

In Figs.\ref{fig2}(a) and \ref{fig2}(b) the results have been obtained in the absence of the RWA where both the interactions have been considered in the system Hamiltonian, i.e., $\epsilon_b=1,\epsilon_t=1$ while in Figs.\ref{fig2}(c) and \ref{fig2}(d) the results have been obtained in the presence of the RWA when just the BS interaction has been considered in the system Hamiltonian i.e., $(\epsilon_b=1,\epsilon_t=0)$. The reason why the TMS interaction can be ignored in the red detuned regime of $\Delta=\omega_{m}$ is that it oscillates at the frequency of $2\omega_{m}$ in the interaction picture while the BS interaction has no oscillation \cite{Aspelmeyer}.

In Fig.~(\ref{fig2}) the parameters have been chosen so that the system is in the normal mode splitting regime because $g > \kappa$. Besides, in Fig.~\ref{fig2}(a) where the the RWA has not been considered the Stokes amplitude $D_S$ (the red dashed curve) is non-zero while as is seen from Fig.~\ref{fig2}(c) in the presence of the RWA it is zero. Based on Eqs.~(\ref{Gaa(w)}) and (\ref{Xaad}) the Stokes amplitude is dependent on the effective modulation parameter $\lambda_a(\omega)$ which gets zero for $g_t=0$ (see Eq.~(\ref{lambdaa})). It means that a non-zero Stokes amplitude in the red-detuned regime is predictable in the model of the standard OMS without considering the RWA and its value increases by increasing the effective optomechanical coupling. 

\begin{figure}
	\centering
	\includegraphics[width=4.25cm]{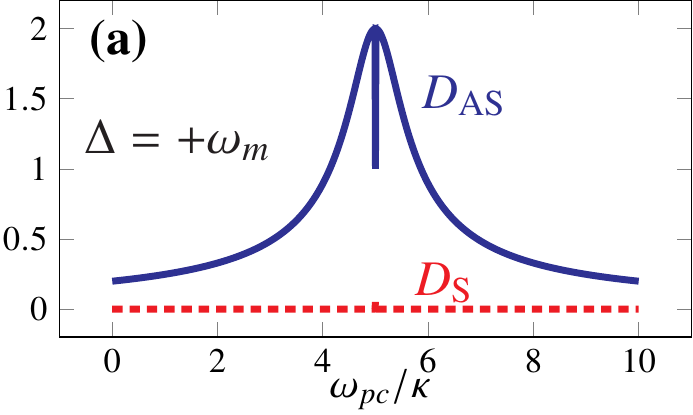}
	\includegraphics[width=4.25cm]{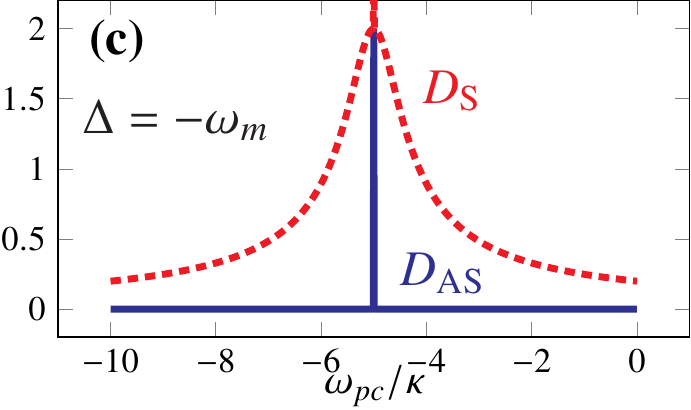}
	\includegraphics[width=4.25cm]{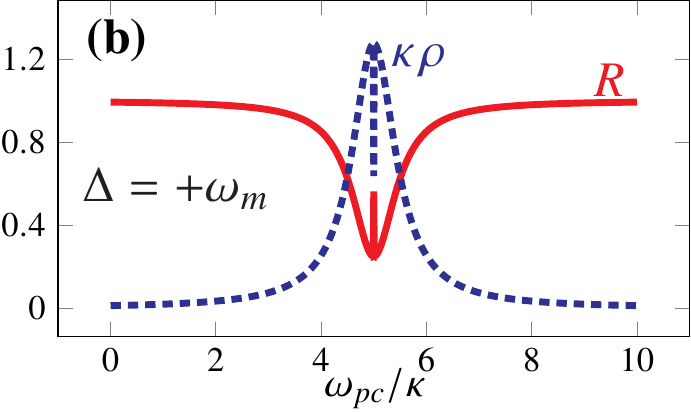}
	\includegraphics[width=4.25cm]{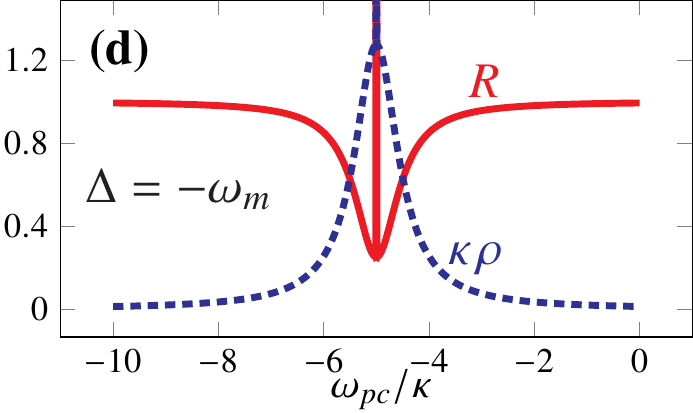}
	\caption{
		(Color online) (a),(c) The optical amplitudes of the anti-Stokes $D_{AS}$ (blue solid curve) and the Stokes $D_{S}$ (red dashed curve) sidebands and (b), (d) the normalized spectral density function $\kappa\rho$ (blue dashed curve) and the reflection coefficient $R$ (red solid curve) versus the normalized frequency $\omega_{pc}/\kappa$. Here, all the results have been obtained in the absence of the RWA $(\epsilon_b=1,\epsilon_t=1)$  with the optomechanical coupling of $g=0.005\kappa$. In (a) and (b) the system is in the red-detuned regime while in (c) and (d) it is in the blue-detuned regime. The other parameters are the same as those of Fig.(\ref{fig2}).}
	\label{fig4}
\end{figure}

The important point is that, there are two resonances at the normal frequencies of the OMS because of the coupling between the optical and mechanical modes while there is an anti-resonance at $\omega_{pc}\approx\omega_{m}$ which is equivalent to $\omega_{p}\approx\omega_0$ where the anti-Stokes amplitude gets zero. The phenomenon of the anti-resonance \cite{Belbasi,Joe} occurs in every dynamical system (no matter being quantum or classical) consisting of two or several coupled oscillators. Specifically, in the case of two coupled oscillators, like the OMS, there is one anti-resonance frequency just for the oscillator which is directly driven by the external source where its amplitude of oscillation goes to zero. That is why in the OMS the amplitude of the optical field which is the oscillator that is driven by the external source goes to zero at the anti-resonance frequency. In the next subsection, where we study the linear response of the mechanical mode, it is shown that at the anti-resonance frequency the mechanical mode oscillates with a non-zero amplitude [see Fig.~(\ref{fig6})].

Furthermore, the anti-resonance phenomenon is manifested in the limit where the damping rate of the second oscillator is negligible in comparison to that of the first one which is directly driven by the external source. Under this condition it can be shown \cite{Belbasi,Joe} that the phase of the first oscillator suffers a sudden change and becomes out of phase with the second one so that the motion of the first oscillator is quenched effectively by the second one. In this way, the amplitude of the first oscillator goes to zero when the source frequency gets near to the anti-resonance frequency while it increases suddenly as the source frequency increases above the anti-resonance. That is why the slope of the anti-Stokes amplitude in Fig.~(\ref{fig2}) suffers a sudden change, as in the present OMS the condition of $\gamma\ll\kappa$ is satisfied and all the features of the anti-resonance are completely observable.

On the other hand, as is seen from Eqs.~(\ref{Gaadagwp}) and (\ref{LRS empcav}) which give, respectively, the Green's function and the linear response of a standard optical cavity with fixed mirrors, there is just one resonance at $\omega_{p}=\omega_0$ due to the presence of the single mode of the system (optical cavity mode). It means that, the anti-resonance of the OMS occurs at the resonance frequency of a similar cavity with fixed mirrors. This is the well-known phenomenon of the OMIT which leads to the appearance of a transparency window around the resonance frequency of the cavity due to the normal mode splitting and the existence of an anti-resonance at the resonance frequency of the cavity. 

In Figs.~\ref{fig2}(b) and \ref{fig2}(d) the cavity spectral function (blue dashed curve) and the power reflection coefficient (red solid curve) have been plotted versus the normalized frequency $\omega_{pc}/\kappa$. As has been already stated the OMIT occurs at the cavity resonance where the cavity spectral function goes to zero and, consequently, the power reflection goes to unity. As is seen from Figs.~\ref{fig2}(b) and \ref{fig2}(d), at each normal frequency the spectral function reaches a peak and, consequently, the power coefficient reduces to a minimum. On the other hand, at the cavity resonance where $\omega_{pc}=\omega_{m}=5\kappa$, i.e., at $\omega_{p}=\omega_0$ the spectral function vanishes and the reflection coefficient becomes unity and that is why the OMIT occurs there. Besides, since the Stokes amplitude is vanished in the presence of the RWA, the peaks of Figs.~\ref{fig2}(c) and \ref{fig2}(d) are more symmetrical.

In Figs.~(\ref{fig3}) and (\ref{fig4}) the optical amplitudes of the Stokes and anti-Stokes sidebands as well as the cavity spectral function and power reflection coefficient have been plotted versus the normalized frequency $\omega_{pc}/\kappa$ in the absence of the RWA where both the BS and TMS interactions have been considered in the Hamiltonian ($\epsilon_b=1, \epsilon_t=1$) in two different regimes of red $(\Delta=+\omega_{m})$ and blue $(\Delta=-\omega_{m})$ detuning for two values of the effective optomechanical coupling $g=0.5\kappa$ (Fig.~\ref{fig3}) and $g=0.005\kappa$ (Fig.\ref{fig4}). Here, the results based on the presence of the RWA have not been shown because they are similar to the present results in the absence of the RWA due to the smallness of the optomechanical coupling in comparison to $\kappa$.

Here, in the red-detuning regime where $\omega_{c}=\omega_0-\omega_{m}$ (Figs.~\ref{fig3}(a) and \ref{fig4}(a)), tuning the probe frequency around the cavity resonance, i.e., $\omega_{p}\approx\omega_0$ is equivalent to $\omega_{pc}\approx+\omega_{m}$. So, in the red-detuning regime the normalized Stokes and anti-Stokes amplitudes are, respectively, $D_S=\kappa|\tilde G^R_{aa}(-\omega_{pc}/\kappa)|$ and $D_{AS}=\kappa|\tilde G^R_{aa^\dag}(\omega_{pc}/\kappa)|$ based on Eqs.~(\ref{DAS},\ref{DS}) while in the blue-detuning regime where $\omega_{c}=\omega_0+\omega_{m}$ (Figs.~\ref{fig3}(c) and \ref{fig4}(c)), the condition of the probe frequency being tuned around the cavity resonance, i.e., $\omega_{p}\approx\omega_0$ is equivalent to $\omega_{pc}\approx-\omega_{m}$. So, in the blue-detuning regime, the Stokes and anti-Stokes amplitudes are, respectively, $D_S=\kappa|\tilde G^R_{aa^\dag}(\omega_{pc}/\kappa)|$ and $D_{AS}=\kappa|\tilde G^R_{aa}(-\omega_{pc}/\kappa)|$  based on Eqs.~(\ref{DAS},\ref{DS}). That is why the results of Figs.~\ref{fig3}(a) and \ref{fig3}(b) have been demonstrated in the positive frequencies around $+\omega_{m}$ while those of Figs.~\ref{fig3}(c) and \ref{fig3}(d) have been shown in the negative frequencies around $-\omega_{m}$ where the peaks corresponding to the normal modes are appeared.

\begin{figure}
	\centering
	\includegraphics[width=8.4cm]{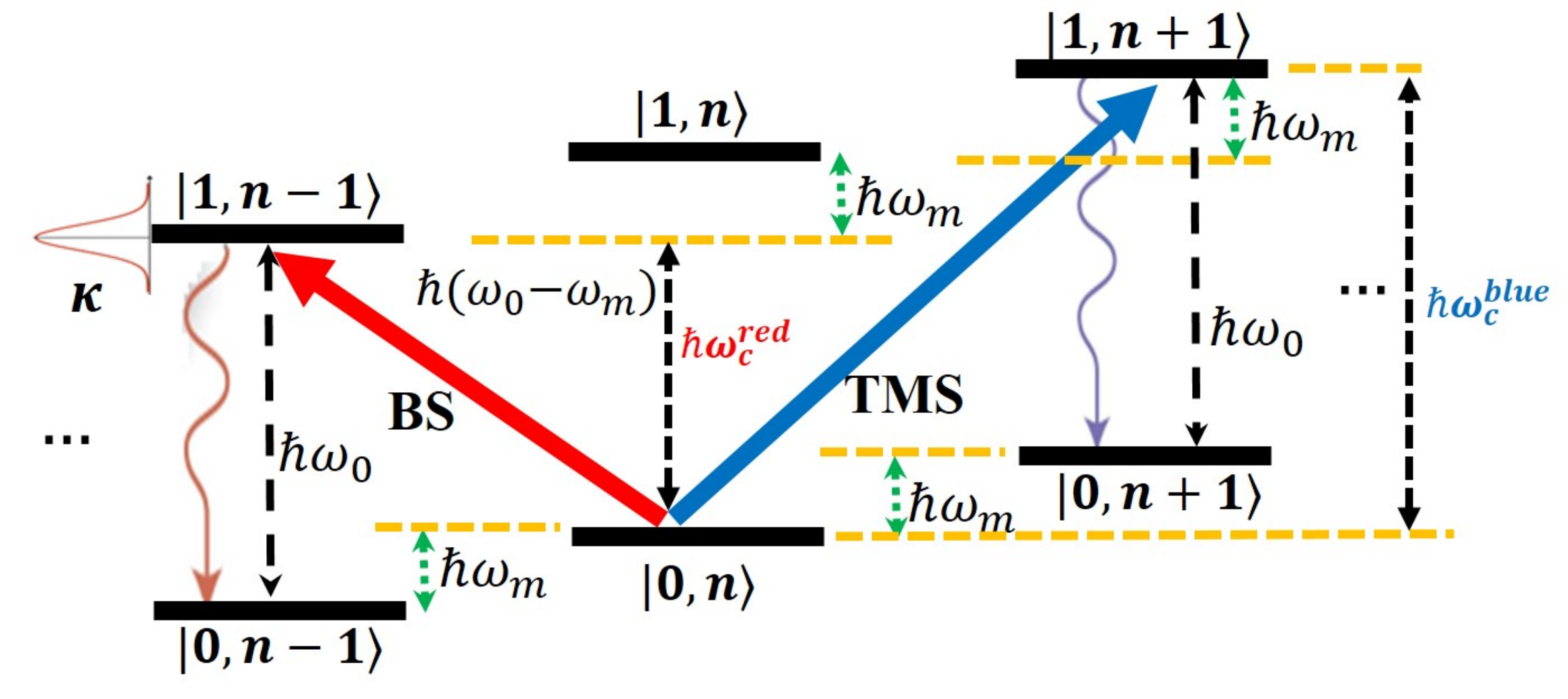}
	\caption{
		(Color online) The energy-level diagram of a standard OMS. Here, 0 and 1 correspond to the optical mode while $n$ corresponds to the mechanical mode excitations. In the red-detuned regime where $\omega_{c}^{red}=\omega_0 - \omega_{m}$, the domination of the BS interaction leads to the amplification of the anti-Stokes sideband while in the blue-detuned regime where $\omega_{c}^{blue}=\omega_0 + \omega_{m}$, the domination of the TMS interaction leads to the amplification of the Stokes sideband.}
	\label{fig5}
\end{figure}

The important point that should be noted is that in the red-detuning regime $(\Delta=\omega_{m})$ the anti-Stokes sideband is amplified and the Stokes sideband is suppressed while in  the blue-detuning regime $(\Delta=-\omega_{m})$ the Stokes sideband is amplified and the anti-Stokes sideband is suppressed when the probe frequency is tuned around the cavity resonance. This result is compatible with the analysis based on the energy-level transitions as has been shown in Fig.~(\ref{fig5}). 

Based on the Raman-scattering picture \cite{Aspelmeyer,Wilson} which is valid in the weak-coupling regime ($g\ll\kappa$) like the results obtained in Fig.(\ref{fig4}), in the red-detuned regime of $\Delta=+\omega_m$ the coupling laser is approximately on resonance with the transition $|0,n\rangle \to |1,n-1\rangle$ while in the blue-detuned regime of $\Delta=-\omega_m$ it is on resonance with the transition $|0,n\rangle \to |1,n+1\rangle$ where $n$ corresponds to the number of the mechanical mode excitations [see Fig.(\ref{fig5})]. In the former where the BS interaction is dominant the input photon absorbs a phonon from the mechanical oscillator and is reflected blue-shifted by $\omega_{m}$ which leads to the amplification of the anti-Stokes sideband while in the latter  where the TMS interaction is dominant the input photon gives a phonon to the mechanical oscillator and is reflected red-shifted by $\omega_{m}$ which leads to the amplification of the Stokes sideband.

The other point is that the weaker is the effective optomechanical coupling, the nearer will be the normal modes to each other and the narrower becomes the transparency window (compare Figs.~\ref{fig2}(a), \ref{fig3}(a), and \ref{fig4}(a)) so that for very low values of $g$ in comparison to $\kappa$ (like that demonstrated in Fig.\ref{fig4}) the splitting between the two modes goes to zero and the two peaks corresponding to the two normal modes approximately merge into each other so that just one peak can be  practically observed at $\omega_{pc}=\pm\omega_{m}$ as has been shown in Fig.(\ref{fig4}).
	
Furthermore, Figs.~\ref{fig3}(b) and \ref{fig3}(d) confirm that the reflection coefficient minimizes at the normal mode frequencies where the cavity spectral function maximizes while at the cavity resonance frequency where the anti-resonance occurs the cavity spectral function reduces to zero and, consequently, the reflection coefficient becomes unity. The same phenomenon occurs for $g=0.005\kappa$ as has been shown in Figs.~\ref{fig4}(b) and \ref{fig4}(d) with the difference that the width of transparency window goes to zero so that there is practically one peak of spectral function around the cavity resonance. For such a low value of optomechanical coupling the peak of the reflection coefficient at the anti-resonance point is very sharp because of the narrowness of the transparency window.

\subsection{Mechanical Response of the OMS}
In order to investigate the linear response of the mechanical mode fluctuation $\delta\hat b(t)$ to the external time-dependent perturbation of Eq.~(\ref{Vt}) it is enough to substitute Eq.~(\ref{VIt}) into Eq.~(\ref{EAt}) with $\hat u=\delta b$ which leads to the following equation
\begin{eqnarray}\label{LRT bt}
\langle\delta\hat b(t)\rangle=\langle\delta\hat b\rangle_0+\zeta\int_{-\infty}^{+\infty} dt^{\prime} G_{ba}^R(t-t^{\prime}) &&e^{i\omega_{pc}t^{\prime}}\nonumber\\
+\zeta^\ast\int_{-\infty}^{+\infty} dt^{\prime} G_{ba^\dag}^R(t-t^{\prime}) e^{-i\omega_{pc}t^{\prime}},
\end{eqnarray}
where $\langle\delta\hat b\rangle_0=0$ is the steady-state mean value of the operator $\delta\hat b$ in the absence of the time-dependent perturbation and
\begin{subequations}
	\begin{eqnarray}
	G_{ba}^R(t)=-i\theta(t)\langle [\delta\hat b(t),\delta\hat a(0)]\rangle_0,\label{Gba}\\
	G_{ba^\dag}^R(t)=-i\theta(t)\langle [\delta\hat b(t),\delta\hat a^\dag(0)]\label{Gbad}\rangle_0,
	\end{eqnarray}
\end{subequations}
are the retarded Green's functions corresponding to the mechanical mode in which $\delta\hat b(t)$ is in the interaction picture, i.e., it satisfies the QLEs corresponding to the Hamiltonian $\hat H_0$ given by the Eq.(\ref{H0}). It is easy to show that Eq.(\ref{LRT bt}) can be written as
\begin{equation}\label{mbt}
\langle\delta\hat b(t)\rangle=\zeta^\ast \tilde G_{ba^\dag}^R(\omega_{pc}) e^{-i\omega_{pc}t}+\zeta \tilde G_{ba}^R(-\omega_{pc}) e^{i\omega_{pc}t},
\end{equation}
where $\tilde G_{ba^\dag}^R(\omega_{pc})$ and $\tilde G_{ba}^R(-\omega_{pc})$ are the Fourier transforms of the Green's functions of the mechanical mode at $\omega=\pm\omega_{pc}$.

Since $\langle\hat b(t)\rangle=\beta+\langle\delta\hat b(t)\rangle$ with $\beta\approx g_0\alpha^2/(\omega_{m}-i\gamma_m/2)$ being the mechanical mean-field, the response of the mechanical field $\hat b(t)$ to the time-dependent perturbation is obtained as
\begin{equation}\label{mbt2}
\langle\hat b(t)\rangle=\beta +\zeta^\ast \tilde G_{ba^\dag}^R(\omega_{pc}) e^{-i\omega_{pc}t}+\zeta \tilde G_{ba}^R(-\omega_{pc}) e^{-i\omega_{pc}t}.
\end{equation}

The Green's functions of the mechanical mode can be easily obtained from Eqs.(\ref{GAdw}) and (\ref{GAw}) in the Fourier space as
\begin{subequations}
	\begin{eqnarray}
	&& \tilde G^R_{ba^\dag}(\omega)=- i\chi_{ba} (\omega),\label{Gbad(w)}\\
	&& \tilde G^R_{ba}(\omega)=+ i\chi_{ba^\dag} (\omega),\label{Gba(w)}
	\end{eqnarray}
\end{subequations}
where the susceptibility elements of the mechanical mode can be calculated by Eqs.~(\ref{bwso}-\ref{bdagwso}) and Eqs.~(\ref{asolveOMS1 a}-\ref{asolveOMS1 b}) as
\begin{subequations}
	\begin{eqnarray}
	&& \chi_{ba}(\omega)=-i\chi_{-m}(\omega)\Big(g_b \chi_{aa}(\omega)+g_t \chi_{aa^\dag}^{\ast}(-\omega)\Big),\label{Xba(w)}\\
	&& \chi_{ba^\dag}(\omega)=-i\chi_{-m}(\omega)\Big(g_b\chi_{aa^\dag}(\omega)+g_t \chi_{aa}^{\ast}(-\omega)\Big),\label{Xbadag(w)}
	\end{eqnarray}
\end{subequations}
in which $\chi_{-m}$, $\chi_{aa}$, and $\chi_{aa^\dag}$ are, respectively, given by Eqs.~(\ref{Xpmm}), (\ref{Xaa}) and (\ref{Xaad}). 

\begin{figure}
	\centering
	\includegraphics[width=4.25cm]{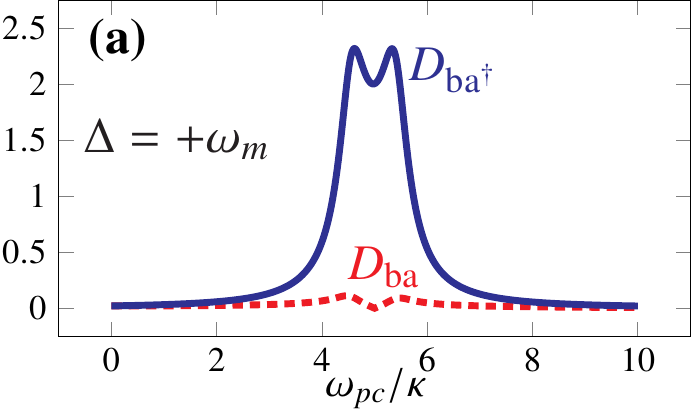}
	\includegraphics[width=4.25cm]{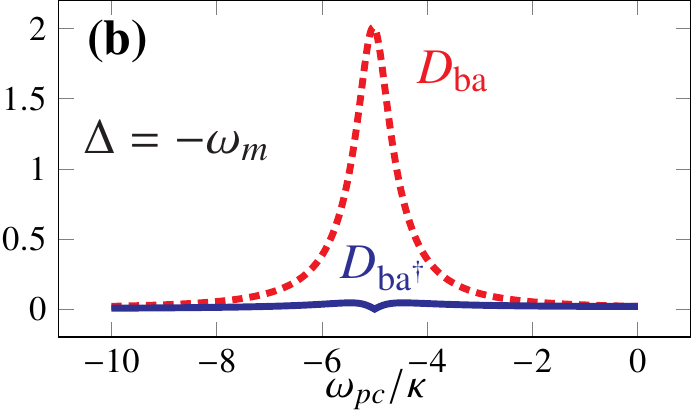}
	\caption{
		(Color online) The normalized mechanical amplitudes $D_{ba^\dag}=\kappa|\tilde G^{R}_{ba^\dag}(\omega_{pc}/\kappa)|$  (blue solid curve) and $D_{ba}=\kappa|\tilde G^{R}_{ba}(-\omega_{pc}/\kappa)|$ (red dashed curve) versus the normalized frequency $\omega_{pc}/k$. Here, all the results have been obtained in the absence of the RWA $(\epsilon_b=1,\epsilon_t=1)$  with the optomechanical coupling of $g=0.5\kappa$. In (a) and (b), it is assumed that the system operates, respectively, in the red- and blue-detuned regimes. The other parameters are the same as those of Fig.(\ref{fig2}).}
	\label{fig6}
\end{figure}

In Fig.~(\ref{fig6}) the normalized amplitudes $D_{ba^\dag}=\kappa|\tilde G^{R}_{ba^\dag}(\omega_{pc}/\kappa)|$ indicated by the blue solid curve and $D_{ba}=\kappa|\tilde G^{R}_{ba}(-\omega_{pc}/\kappa)|$ indicated by the red dashed curve have been plotted versus the normalized frequency $\omega_{pc}/\kappa$ in the red-detuned [Fig.~\ref{fig6}(a)] and blue-detuned [Fig.~\ref{fig6}(b)] regimes when the optomechanical coupling has been considered $g=0.5\kappa$ like that of Fig.~(\ref{fig3}).

Fig.~\ref{fig6}(a) and Fig.~\ref{fig3}(a) depict, respectively, the amplitudes of the oscillations of the optical and the mechanical modes under the same conditions in the red-detuned regime. As is seen, at each normal frequency where the anti-Stokes amplitude of the optical mode reaches a peak, the mechanical mode amplitude $D_{ba^\dag}$ maximizes too. On the other hand, at the anti-resonance frequency of $\omega_{pc}=\omega_{m}$ where the amplitude of optical field gets zero, the mechanical mode amplitude $D_{ba^\dag}$ reaches a local minimum which is approximately as large as each of the peaks. It is exactly the characteristic of the well-known phenomenon of anti-resonance in the coupled oscillators dynamics \cite{Belbasi,Joe}. As has been already explained, at the anti-resonance frequency the amplitude of the directly driven oscillator (the optical mode) goes to zero while the other oscillator (the mechanical mode) oscillates at a finite amplitude.

In the blue-detuned regime, as has been shown in Fig.~\ref{fig6}(b), the amplitude $D_{ba}$ is amplified while the amplitude $D_{ba^\dag}$ is suppressed which is the opposite of the situation in the red-detuned regime. Fig.~\ref{fig6}(b) and Fig.
~\ref{fig3}(b) depict, respectively, the amplitudes of the oscillations of the optical and the mechanical modes under the same conditions in the blue-detuned regime. Again here, at the anti-resonance frequency of $\omega_{pc}=-\omega_{m}$ where the amplitude of optical field (the Stokes amplitude $D_S$) goes to zero, the mechanical mode oscillates with a large amplitude of $D_{ba}$.

\section{Summary, Conclusion and Perspective \label{sec conclusion}}
In this paper, we have presented a \textit{generalized} formulation of the \textit{standard} LRT as raised in condensed matter theory to incorporate the theory of open quantum systems in the Heisenberg picture in order to be able to describe the linear response of an open quantum system to an external time-dependent perturbation. Besides, using the method of Green's function equation of motion, we not only derive the evolutions of the open quantum system Green's functions but also clarify the relations between the system susceptibilities and Green's functions. As a simple example, we have exploited the \textit{generalized} LRT for a single mode open quantum field like the optical field inside a driven standard cavity with fixed mirrors.

Then, as a more important and applicable example for the application of the \textit{generalized} LRT, we have investigated the linear response of a standard OMS to an external time-dependent potential which drives the optical mode of the cavity. We have studied the linear responses of the optical as well as the mechanical modes of a standard linearized OMS to an external time-dependent potential that drives the optical field of the cavity. For this purpose we have introduced a systematic and easy method to derive analytically the equations of motion for Green's functions of an open quantum system like the OMS in the Heisenberg picture. In this way, using the system Green's function one can obtain the optical and mechanical responses of the OMS both in the red- and in the blue-detuning regimes for different values of the optomechanical coupling parameter.

It has been shown that when the optomechanical coupling is comparable or larger than the damping rate of the optical mode, the normal mode splitting phenomenon is observable in the amplitude of the optical field due to the resolvability of the two resonances corresponding to the two normal modes of the OMS . Furthermore, another important phenomenon, the so-called anti-resonance, is observable in the optical mode which leads to a near zero amplitude of the optical field. At the anti-resonance frequency the amplitude of the optical mode (which is the oscillator driven directly by the external source) goes to zero while the mechanical mode oscillates with a finite amplitude. In this way, the explanation of the OMIT phenomenon is easily possible due to the presence of the two resolvable resonances together with an anti-resonance between them which leads to the appearance of a transparency window around the resonance frequency of the cavity.

Furthermore, the \textit{generalized} LRT explains how the anti-Stokes and Stokes sidebands are amplified, respectively, in the red- and blue-detuning regimes. The obtained results are in exact coincidence, especially in the weak-coupling regime, with the Raman-scattering picture which demonstrates how the optical shot noise affects the transition rates between the energy levels of the OMS.

As an interesting outlook, we would like to remind that the Green' function approach of the open quantum systems introduced in Secs.\ref{sec GLRT} and \ref{sec LR OMS} which is based on the method of equations of motion can be exploited for more complex open quantum systems with nonlinearities like those studied in Refs.\cite{greenPRB,greenOMS1,greenPRLclerk,greenOMS2,foroudcrystalentanglement,foroudsynch}. For this purpose, one can use the introduced method of equation of motion series expansion investigated in Ref.\cite{greenPRB} to calculate the open system Green's functions. The other interesting examples are hybrid OMSs consisting of interacting BECs \cite{Dalafi2016,Dalafi2017a,Dalafi2017b,Dalafi2018} and parametrically modulated OMS \cite{optomechanicswithtwophonondriving,aliDCE3,aliDCEBECsqueezing,aliNJP,aliDCE1,aliDCE2,aliAVSbook2020} as well as amplifier/squeezer magnonic system \cite{mehryMagnetometry2020} whose linear response to the external perturbation could be very interesting to be investigated. It should be noted that one of the advantages of such schemes is their capabilities for ultra-precision quantum measurements \cite{aliDCEBECforce,Fani2020}. Interestingly, the quantum limit on the sensitivity of such quantum detectors can be calculated by their linear responses \cite{clerk2003,clerk2004}. Besides, investigation of the linear response for Gaussian continuous variable quantum systems has been very important in quantum information in recent years \cite{mehboudi 2018,mehboudi 2019}.

Furthermore, as we will show in a future paper, the parametrically driven OMS introduced in Refs.~\cite{optomechanicswithtwophonondriving,aliDCE3,aliDCEBECsqueezing,aliDCEBECforce,aliDCE1,aliDCE2,aliNJP} has the potential to manifest negative cavity photon spectral function. This negativity may lead to a new class of OMIT phenomenon in which the power reflection coefficient of the probe laser grows above the unity near the resonance frequency of the cavity. It also leads to the definition of a negative effective temperature for the intracavity photons resulting in the population inversion of a qubit coupled to the parametrically driven OMS. The other interesting work is exploration for Floquet periodic driving systems in non-Markovian regimes as has been recently investigated \cite{shenOptLett18}. On the other hand in the Markovian regime, one can use a perturbative approach to calculate corrections to eigenstates and eigenvalues of the Liouvillian super-operator of the master equation \cite{Li}.

\begin{acknowledgements}
AMF would like to thank the Office of the Vice President for Research of the University of Isfahan for their support. AD gratefully acknowledges support from the Iran National Science Foundation (INSF) under Grant No. 99020597.
\end{acknowledgements}

\appendix
\section{QLEs for a single-mode quantum field}\label{appA}
Based on the theory of open quantum systems \cite{zubairy,Gardiner}, the environment of an open quantum system is modeled as a multi-mode bosonic quantum field in the thermodynamic equilibrium with the Hamiltonian $\hat H_R=\hbar\sum_{k}\omega_k\hat b^\dagger_{k}\hat b_{k}$ which interacts with the system through the interaction Hamiltonian $\hat H_{SR}=\hbar\sum_{k}g_{k}(\hat b^\dagger_{k}\hat a+\hat a^\dag \hat b_{k})$ where $\hat b_{k}$ is the annihilation operator of the mode $k$ of the reservoir and $g_{k}$ is its coupling parameter with the single mode of the system. 

Since the total system consisting of the reservoir and the system can be considered as a closed one, the equations of motion for the degrees of freedom of the system and reservoir are obtained from the Heisenberg equations of motion, i.e., $i\hbar\dot{\hat O}=[\hat O, \hat H]$ for ($\hat O=\hat a$ or $\hat b_k$) which can be written as
\begin{subequations}
	\begin{eqnarray}
	\dot{\hat a}(t)=-i\omega_0\hat a(t)-i\sum_{k}g_k\hat b_k(t),\label{aA}\\
	\dot{\hat b}_k (t)=-i\omega_k\hat b_k(t)-ig_k\hat a(t).\label{bkA}
	\end{eqnarray}
\end{subequations}
It is obvious that the time-evolution of the system variable $a(t)$ can also be obtained from the followin equation
\begin{equation}
\hat a(t)=e^{\frac{i}{\hbar}\hat H (t-t_0)} \hat a(t_0) e^{-\frac{i}{\hbar}\hat H (t-t_0)},\label{atA}\
\end{equation}
which can be considered as the solution to the differential equations of Eqs.(\ref{aA}-\ref{bkA}). It should be reminded that Eq.(\ref{atA}) is nothing except for Eq.(\ref{uIt}) for $\hat u=\hat a$ and $\hat H_0=\hat H$ in which the indext of $I$ has been ignored.

Now, by integrating Eq.(\ref{bkA}) formally and substituting in Eq.(\ref{aA}), the equation of motion for the field operator $\hat a$ is obtained as follows
\begin{equation}\label{aA2}
\dot{\hat a}(t)=-i\omega_0\hat a(t)-\sum_{k} g_{k}^{2}\int_{t_0}^{t}dt' a(t') e^{-i\omega_k (t-t')}+ \hat\xi (t),
\end{equation}
where $\hat{\xi}(t)$ is the quantum noise operator which has been defined as $\hat{\xi}(t)=-i\sum_{k} g_k \hat b_k (t_0) e^{-i\omega_k (t-t_0)}$. Assuming that the modes of the reservoir are closely spaced in frequency, the summation over $k$ can be replaced by an integral which under the Wigner-Weisskopf approximation yields a delta function \cite{zubairy}. In this way, Eq.(\ref{aA2}) in the continuum limit takes the following form
\begin{equation}
\dot{\hat a}(t)=-(i\omega_0+\kappa/2)\hat a(t)+\sqrt{\kappa}\hat a_{in}(t). \label{adottA}\\
\end{equation}
Here, the damping rate of the quantum field, i.e., $ \kappa $, is given by $\kappa=2\pi g^2(\omega_0/c)\sigma(\omega_0)$ in which $\sigma(\omega_0)=V\omega_0^2/\pi^2 c^3$ is the density of the reservoir states with $V$ being the quantization mode volume. Furthermore, the quantum noise operator $\hat a_{in}(t)$ in Eq.(\ref{adottA}) is the continuum limit counterpart of the operator $\hat\xi(t)$ with the following correlation functions \cite{zubairy}
\begin{subequations} \label{correlation}
	\begin{eqnarray}
	&&	\left\langle\hat a^{\dagger}_{in} (t) \hat a_{in}(t') \right\rangle = \bar{n}_{th} \delta (t-t')\\
	&& \left\langle\hat a_{in} (t) \hat a^{\dagger}_{in}(t') \right\rangle = (\bar{n}_{th} +1) \delta (t-t'),
	\end{eqnarray}
\end{subequations}
where $ \bar{n}_{th} = (e^{\hbar \omega_0 / k_B T}-1)^{-1} $ is the mean number of thermal excitations.

\end{document}